\documentclass[a4paper,11pt]{article}
\pdfoutput=1 

\usepackage{jcappub} 

\usepackage[T1]{fontenc} 

\usepackage{bm}
\usepackage{braket}
\newcommand{\ODD}[2]{\frac{\mathrm{d} #1}{\mathrm{d} #2}}

\usepackage{here}

\title{\boldmath Axion clouds may survive the perturbative tidal interaction over the early inspiral phase of black hole binaries}

\author[a]{Takuya Takahashi}
\author[a,b]{and Takahiro Tanaka}

\affiliation[a]{Department of Physics, Kyoto University, Kyoto 606-8502, Japan}
\affiliation[b]{Center for Gravitational Physics, Yukawa Institute for Theoretical Physics, Kyoto University, Kyoto 606-8502, Japan}

\emailAdd{t.takahashi@tap.scphys.kyoto-u.ac.jp}
\emailAdd{t.tanaka@tap.scphys.kyoto-u.ac.jp}

\abstract{Gravitational wave observation has the potential of probing ultralight bosonic fields such as axions. Axions form a cloud around a rotating black hole (BH) by superradiant instability and should affect the gravitational waveform from binary BHs. On the other hand, considering the cloud associated with a BH in a binary system, tidal interaction depletes the cloud in some cases during the inspiral phase. We made an exhaustive study of the cloud depletion numerically in a wide parameter range for equal mass binaries, assuming only the leading quadrupolar tidal perturbation is at work as a first step. Under this assumption, we found that clouds can avoid disappearing due to the tidal effect only when (1) the $l=1$ mode, which refers to the lowest bound state with the azimuthal angular momentum eigenvalue being $+1$, is the fastest growing mode, (2) the binary orbit is counter-rotating relative to the BH spin and (3) the cloud is in the non-relativistic regime.}

\begin{document}
\maketitle
\flushbottom

\section{Introduction}
\label{sec:intro}
The era of gravitational wave (GW) observation has begun with the first direct detection by LIGO and Virgo collaboration~\cite{Abbott:2016blz,TheLIGOScientific:2016qqj}, and now many binary black hole (BH) merger events have been reported~\cite{Abbott:2020niy}. Through the observation of GWs, a variety of scientific discoveries are expected, including physics beyond the Standard Model. As one interesting example of it, we focus on the possibility of detecting ultralight bosonic fields such as axions.

The existence of plenty of axion-like particles over a broad range of mass is predicted by string theory, and this scenario is called ``axiverse''~\cite{Arvanitaki:2009fg}. Such particles are thought to leave observable footprints on cosmology or astrophysics. In particular, we focus on a massive scalar field such as an axion around a rotating BH. The scalar field can be bounded gravitationally and can grow exponentially extracting the energy and the angular momentum from the BH. This is called the superradiant instability~\cite{Brito:2020oca,Dolan:2007mj}. By this instability, a massive scalar field can form a macroscopic condensate around the BH, and we refer to it as an axion cloud. The growth rate becomes large when the Compton wavelength $\lambda_{c}=\hbar/\mu c$ is comparable to the gravitational radius of the BH $r_{g}=GM/c^2$, where $\mu$ is the mass of the scalar field and $M$ is the mass of the BH. Therefore, this phenomenon is relevant to the observation with the axion in the mass range $10^{-20}\sim10^{-10}$ eV for astrophysical BHs. Several methods have been proposed to probe this cloud observationally. For example, detection of GWs emitted from an axion cloud or looking for gaps in the distribution in the spin versus mass plane (Regge plane) of astrophysical BHs~\cite{Arvanitaki:2010sy,Arvanitaki:2014wva,Arvanitaki:2016qwi,Yoshino:2013ofa,Brito:2014wla,Brito:2017zvb}.

Because the GW events observed so far are those originating from binary coalescences, and the number of events is certainly going to increase in the near future, it would be very important to consider the dynamics of axion clouds in binary systems. In Refs.~\cite{Baumann:2018vus,Baumann:2019ztm,Berti:2019wnn}, the dynamics and GW signatures of axion clouds around BHs in binary inspirals were studied. They showed that tidal perturbations from the binary companion induce resonant level transitions of axion clouds. Axions occupying the growing mode are transferred to the decaying mode and this transition depletes the clouds in some cases. For stellar mass BH binaries, the cloud depletion occurs typically in the frequency band to which space-based detectors, such as LISA, are sensitive.

The presence of axion clouds around the BHs in a binary system should affect the waveform especially near the merger-ringdown phase~\cite{Choudhary:2020pxy}. If the clouds survive to the late inspiral or the merger phase, it may be possible to probe the presence of an axion cloud even with ground-based GW detectors. In Refs.~\cite{Baumann:2018vus,Berti:2019wnn}, the cloud depletion was first studied assuming a constant binary orbital frequency.  Later in Ref.~\cite{Baumann:2019ztm}, it was shown that in the evolution of the binary the transition occurs at the resonance frequency instantaneously and almost all axions are transferred to another mode after the binary passes the resonance frequency, if the transition is slow enough to be recognized as adiabatic. In this work, the regime in which the cloud is non-relativistic ($\lambda_{c}\gg r_{g}$) was mainly studied. 
 
When the cloud is relativistic, the growth rate is large, and hence such a system would be more interesting as an observational target. In such cases, the non-relativistic analytic approximation to the eigenfrequency of a quasi-bound state of an axion is not enough, and non-trivial transitions also can occur. Furthermore, considering the self-interaction of an axion, the energy extraction from the BH by superradiance may terminate before the BH spin drops to the threshold for the superradiance~\cite{Arvanitaki:2010sy,Baryakhtar:2020gao,Omiya:2020vji}. Therefore, it makes sense to investigate cloud depletion in a wide parameter region including the relativistic regime.

This paper is the first paper of our series of papers on this subject. 
As a first step, here we restrict our consideration only to the quadrupolar tidal perturbation and assume that the transition is adiabatic. Under this restriction, we investigate which transitions occur numerically including axion's higher angular momentum mode. Also, for equal mass binaries, we examine whether the clouds disappear or not during the early inspiral phase by comparing the lifetime of the binary with the decay rate of axions in the state of the transition destination. Under these assumptions, we clarify the situation where the clouds do not disappear during the early inspiral phase. The result will give a foundation to an exhaustive study of perturbative tidal effect and would be useful also in selecting the setup that requires more detailed investigation, including numerical simulation of the evolution of an axion cloud in a binary system.
 
This paper is organized as follows. In section~\ref{Review}, we give a brief review of the evolution of axion clouds around BHs and how transition and depletion occur in binary inspirals. In section~\ref{main}, we present the numerical results that show which transition first occurs depending on the parameters
and whether the clouds disappear or not. We give a conclusion and discuss the implication of our results in section~\ref{conclusion}, including the scope of the further extension. In the rest of this paper, we use the unit with $G=c=\hbar=1$ unit.

\section{A brief review of Axion Clouds in Binary Systems}\label{Review}

In this section, we briefly review how ultralight scalar fields such as axions form clouds around rotating BHs like hydrogen atoms by the superradiant instability, and how resonant level transitions occur when a BH with an axion cloud is in an inspiraling binary.  

\subsection{Axion clouds around rotating black holes}

Rotating BHs are described by the Kerr spacetime. Consider a real scalar field (axion) $\phi$ of mass $\mu$ on the Kerr metric $g_{\mu\nu}$, which is characterized by the mass $M$ and the angular momentum $J=aM$ of the BH. The equation of motion of a scalar field is
 \begin{equation}\label{eom}
 \left(g^{\mu\nu}\nabla_{\mu}\nabla_{\nu}-\mu^2\right)\phi(x)=0 \quad .
 \end{equation}
Solutions of this equation can be obtained in the form of~\cite{Brill:1972}
 \begin{equation}
 \phi(x)={\rm{Re}}\left[e^{-i(\omega t-m\varphi)}S_{lm\omega}(\theta)R_{lm\omega}(r)\right] \quad ,
 \end{equation}
where $l,m$ are the labels of the spheroidal harmonics $e^{im\varphi}S_{lm\omega}(\theta)$.  Owing to the mass of the scalar field, this system admits quasi-bound state solutions that vanish at the spatial infinity and are purely ingoing waves at the BH horizon. Under these boundary conditions, there exist solutions corresponding to infinitely many, discrete eigenfrequencies, which are complex because of the energy transfer through the horizon. Thus, we denote eigenfrequencies as
 \begin{equation}
 \omega_{nlm}=(\omega_{R})_{nlm}+i(\omega_{I})_{nlm} \quad ,
 \end{equation}
 where $n>0$ denotes the order of the overtone. Without loss of generality, we can assume $(\omega_R)_{nlm}>0$. 
 For modes satisfying the superradiance conditions
 \footnote{Strictly speaking, for complex $\omega$, the superradiance condition becomes
 $|\omega|^2/\omega_{R}<m\Omega_{H}$~\cite{Dolan:2007mj},
 but the imaginary part of the eigenfrequencies of bound states is always very small compared to the real part.}
 \begin{equation}\label{SRcond}
  \omega_{R}<m\Omega_{H} \quad ,
 \end{equation}
and $\omega_{I}$ becomes positive, where $\Omega_{H}:=a/2Mr_{+}$ is the angular velocity of the BH horizon and $r_{+}=M+\sqrt{M^2-a^2}$ is the horizon radius. This positive imaginary part of the frequency means that a gravitationally bound scalar field continues to grow to form a heavy cloud extracting the energy and the angular momentum from the BH.
 
 In the non-relativistic limit, it is appropriate to start with the ansatz 
 \begin{equation}\label{NRphi}
 \phi(t,\bm{r})=\frac{1}{\sqrt{2\mu}}\left[\psi(t,\bm{r})e^{-i\mu t}+{\rm{c.c.}}\right] \quad ,
 \end{equation}
 and $\psi(t,\bm{r})$ is a complex scalar function that slowly varies on time scales much longer than $\mu^{-1}$. Substituting Eq.~(\ref{NRphi}) into Eq.~(\ref{eom}), 
 and keeping only the leading order terms in $r^{-1}$, we obtain
 \begin{equation}\label{NReom}
 i\frac{\partial}{\partial t}\psi=\left(-\frac{1}{2\mu}\nabla^2-\frac{M\mu}{r}\right)\psi \quad .
 \end{equation}
 This equation is formally equivalent to the Schr${\rm{\ddot{o}}}$dinger equation for a hydrogen atom. For smaller $M\mu$, the radius of the cloud is larger, and hence the effect of the presence of the horizon is weaker.  In the non-relativistic limit
 the eigenfunctions $\psi_{nlm}$ are reduced to the ones for a hydrogen atom, and the eigenfrequencies can be given by 
 \begin{equation}\label{NRor}
 (\omega_{R})_{nlm}=\mu\left(1-\frac{(M\mu)^2}{2n^2}+\Delta \omega_{nlm}\right) \quad ,
 \end{equation}
 where $\Delta \omega_{nlm}$ represents higher-order corrections in $M\mu$, whose approximate expression is given in Ref.~\cite{Baumann:2018vus}. For example, this correction breaks the degeneracies between modes with the same $n$ and $l$ but with different $m$ by an order of $(M\mu)^5$. Note that the labels of eigenstates that we can take are restricted to $n>0,l\leq n-1,|m|\leq l$. Approximate formulas for the imaginary part $(\omega_{I})_{nlm}$ are given by Detweiler~\cite{Detweiler:1980uk}.
 
\subsection{Tidal interaction effect from the binary companion}
Following Ref.~\cite{Baumann:2019ztm}, we review what happens when a BH with an axion cloud forms a binary system. The tidal perturbation from the binary companion can be described by an additional potential $V_{*}(t)$ in Eq.~(\ref{NReom}).  We denote the eigenstates of the axion cloud as $\ket{i}$, and a general bound state can be written by a superposition of eigenstates, i.e.,  $\displaystyle \ket{\psi(t)}=\sum_{i}c_{i}(t)\ket{i}$. As in the case of quantum mechanics, $|c_{i}(t)|^2$ represents the occupation number, and its time evolution is given by
\begin{equation}
i\ODD{c_{i}}{t}=\sum_{j}\mathcal{H}_{ij}(t)c_{j}=\sum_{j}\left[\omega_i\delta_{ij}+V_{ij}(t)\right]c_{j} \quad ,
\end{equation}
where $V_{ij}=\braket{i|V_{*}|j}$ encapsulates the effect of tidal perturbation from the binary companion. For simplicity, let us assume that the binary orbit is of a large radius, quasi-circular and on the plane perpendicular to the BH spin. We can write the perturbation as
\begin{equation}
V_{ij}=\eta_{ij}\exp\left(-i\Delta m_{ij}\int_0^t dt'\Omega(t')\right) \quad ,
\end{equation}
where $\eta_{ij}$ represents the amplitude of the perturbation, $\Delta m_{ij}=m_{i}-m_{j}$ and $\Omega(t)$ is the orbital angular velocity. There are selection rules, i.e., the conditions that $\eta_{ij}$ does not vanish. We summarize the selection rules with the explicit expression of tidal perturbation in appendix~\ref{AppTidal}. If the time variation of orbital angular velocity is sufficiently slow, as is well known from the time-dependent perturbation theory in quantum mechanics, a resonant level transition occurs when the orbital angular velocity coincides with the gap of phase velocities of two levels. Thus, we define the resonance frequency as
\begin{equation}
\pm\Omega_{\mathrm{res}}=\frac{\Delta E_{ij}}{\Delta m_{ij}} \quad ,
\end{equation}
where $\Delta E_{ij}=(\omega_{R})_i-(\omega_{R})_j$ and the upper (lower) sign denotes the case of co-rotating (counter-rotating) orbits. In particular, if the time dependence of the orbital angular velocity is approximated by a linear function as
\begin{equation}
\Omega(t)=\Omega_{\mathrm{res}}+\gamma_{ij} t \quad ,
\end{equation}
the problem to solve is known as the Landau-Zener transition. For example, consider a two-level system with an initial condition with $c_1(-\infty)=1$ and $c_2(-\infty)=0$. A resonant transition occurs at around $t=0$ (resonance frequency), and the occupation number at a late time is given by
\begin{equation}
|c_1(+\infty)|^2=e^{-2\pi z}\ , \quad  |c_2(+\infty)|^2=1-e^{-2\pi z} \quad ,
\end{equation}
where $z\equiv\eta_{ij}^2/|\Delta m_{ij}|\gamma_{ij}$. Intuitively, this quantity $z$ roughly measures whether the transition timescale $\Delta t\sim2\eta/\gamma$ is long enough in passing thorough the resonance band $\Delta\Omega\sim2\eta$. If $z\gg1$, the transition is adiabatic and almost all axions are transferred to the other mode.

For the inspiral phase of a binary, $\eta_{ij}$ and $\gamma_{ij}$ are given by
\begin{align}
\frac{\eta_{ij}}{\Omega_{\mathrm{res}}}&\simeq 0.3\left(\frac{R_{ij}}{0.3}\right)\frac{q}{1+q} \quad , \\
\frac{\gamma_{ij}}{\Omega^2_{\mathrm{res}}}&\simeq  8.0\times 10^{-6}\frac{q}{(1+q)^{1/3}}\left(\frac{M}{30M_{\odot}}\right)^{5/3}\left(\frac{\Omega_{\mathrm{res}}}{1\mathrm{Hz}}\right)^{5/3} \quad ,
\end{align}
where we have defined the mass ratio $q\equiv M_*/M$ with $M_*$ being the mass of the binary companion. $R_{ij}$ characterizes the overlap of different two bound states and typically $R_{ij}\lesssim 0.3$. For small $M\mu$, the fastest growing mode is $\ket{211}$ and the axion cloud is dominated by this mode. Since the orbital angular velocity gradually increases, the transition occurs first between the levels with the lowest resonance frequencies. From the selection rules, only transitions with $\Delta l=2p~(p\in \mathbb{Z})$ and $|\Delta m|=2$ are possible for equatorial orbits. In this case, the growing mode $\ket{211}$ makes a transition to a decaying mode $\ket{21-1}$ ($\ket{31-1}$) for co-rotating (counter-rotating) orbits. The resonance frequencies can be calculated by Eq.~(\ref{NRor}), and $\Omega_{\mathrm{res}}\simeq 2.3\times 10^{-4} \mathrm{Hz}~(30M_{\odot}/M)(M\mu/0.1)^7$ ($0.24 \mathrm{Hz}~(30M_{\odot}/M)(M\mu/0.1)^3$). Therefore, if $q=\mathcal{O}(1)$, $z\gg1$ is satisfied by inspiral orbits and transitions should be adiabatic. As a result, the axion clouds disappear in the early inspiral phase.

\section{Level Transitions and Cloud Depletion}\label{main}
In section~\ref{Review}, we mostly discussed the case $M\mu\ll1$. However, we would be more interested in the case $M\mu\gtrsim0.1$ where the growth rate of clouds becomes large. In such parameter regions, the non-relativistic approximation to the eigenfrequencies is not good enough, and we need numerical calculation in order to understand to which modes the axions occupying the growing mode will be transferred. In addition, since $m\leq l$ and $\mathrm{Re}[\omega^2]$ for bound states satisfying $\mathrm{Re}[\omega^2]<\mu^2$ cannot be much smaller than $\mu^2$, lower $l$ modes can no longer satisfy the superradiance condition~(\ref{SRcond}) for a large value of $\mu$, and a higher-$l$ mode becomes the fastest growing mode. For higher-$l$ modes, the transitions allowed by the selection rules become more diverse, involving more variety of modes in the dominant transition. To understand the fate of axion clouds in binary systems, we first exhaustively study which transitions actually occur. Then, we examine whether or not there is a sufficiently long time for the clouds to disappear before the binary coalescence.

\subsection{Transition map}
\begin{figure}[tb]
\centering 
\includegraphics[width=.45\textwidth]{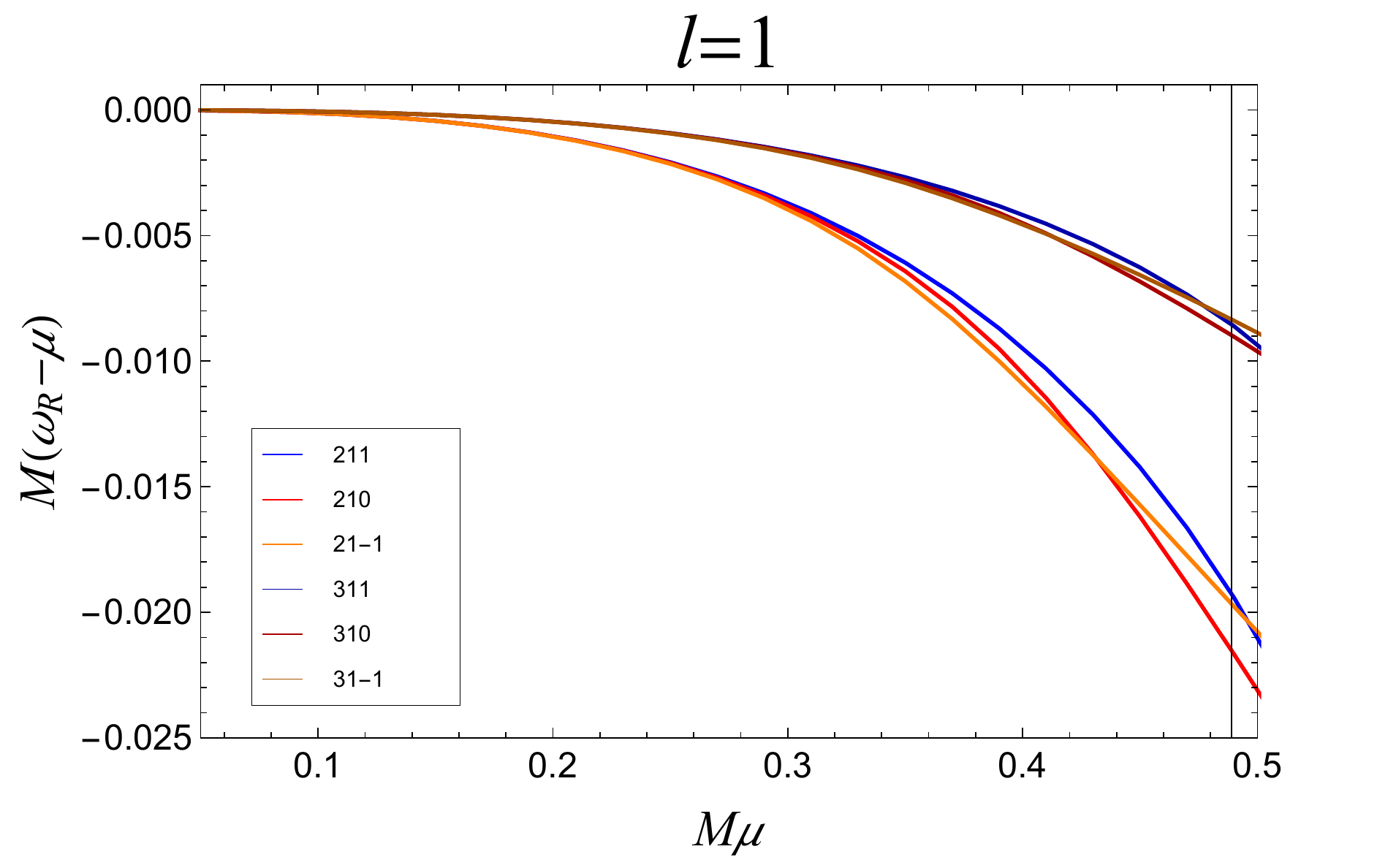}
\hfill
\includegraphics[width=.45\textwidth]{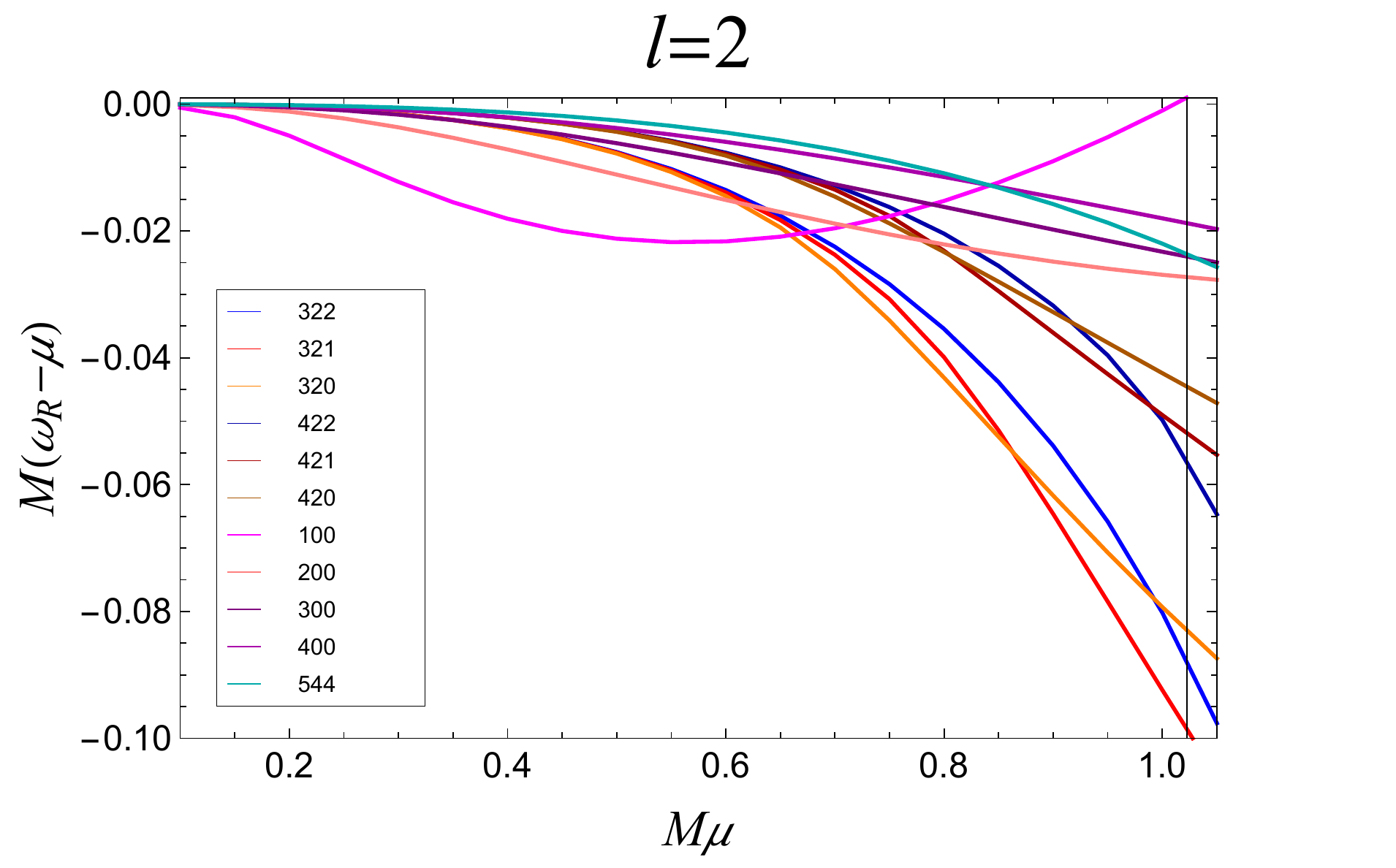} \\
\includegraphics[width=.45\textwidth]{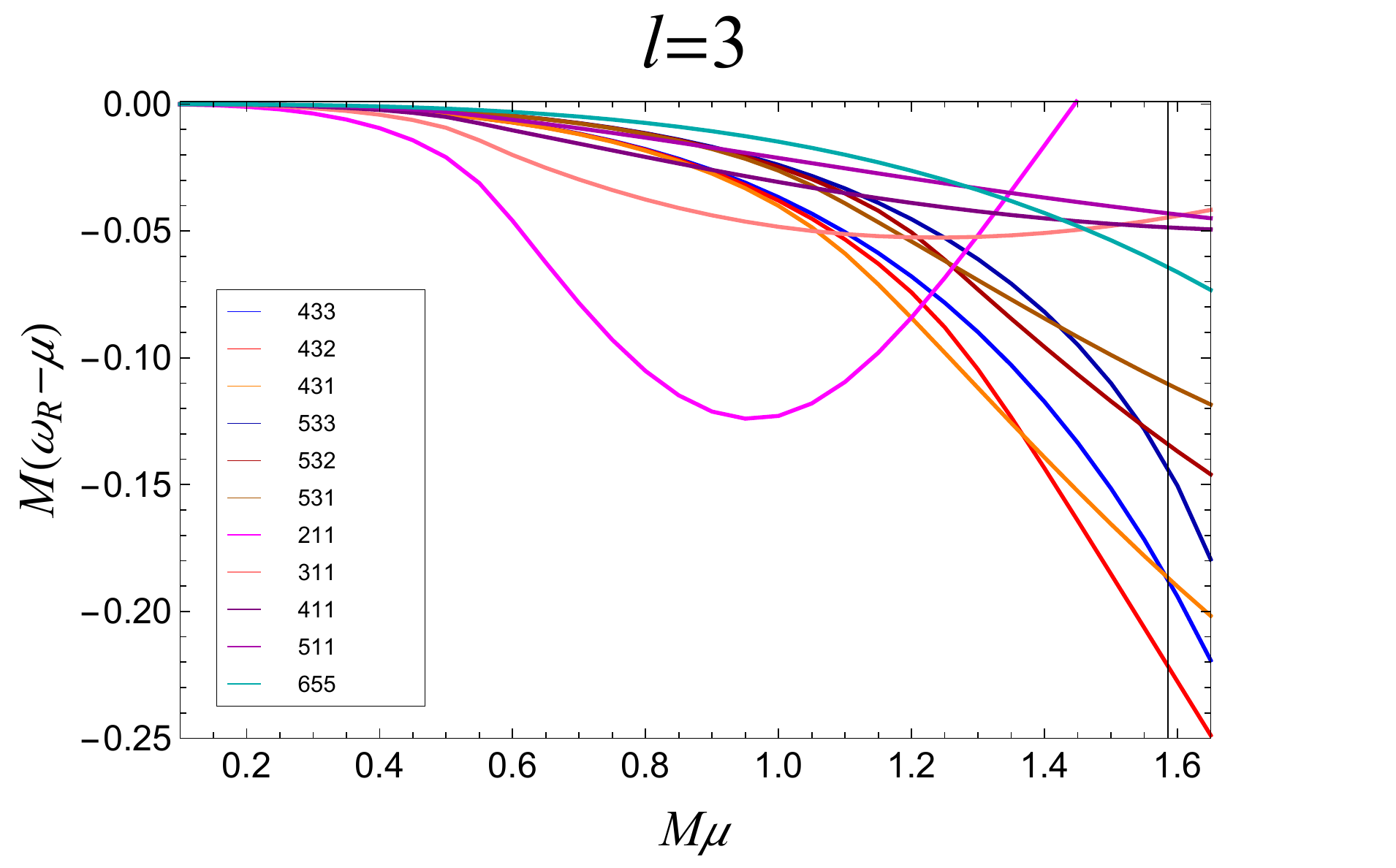} 
\caption{\label{figE} Energy spectra of bound states with $l=1,2$ and $3$ relevant in discussing the transitions for $a/M$=0.998. Considering the quadrupolar tidal perturbation ($l_{\ast}=2$), the axions in $l=1$ mode involve only the transition to $l=1$ mode by the selection rules, but the axions in $l=2$ and $3$ mode can involve the transition to modes with a lower value of $l$. Here, the vertical lines in the respective panels show the range of $M\mu$ beyond which $\ket{nlm}=\ket{211},\ket{322},\ket{433}$ are not the 
fastest growing modes of interest.}
\end{figure}

We investigate to which mode the first transition occurs depending on the binary parameters when $l=1, 2$ or $3$ is the fastest growing mode.  For $l\geq4$ mode, the growth rate is suppressed and it may not be so interesting in a realistic situation. Which transition first occurs is determined by evaluating the energy gaps between the growing mode of interest and the other modes. We calculate the eigenfrequencies of all relevant modes allowed for the transitions numerically using the continued fraction method (see appendix~\ref{AppCF}). They depend on two parameters $(M\mu,a/M)$. As a realistic maximum value of a spin parameter, we adopt the Thorne limit $a/M=0.998$~\cite{Thorne:1974ve}. We show the energy spectra of relevant modes as a function of $M\mu$ 
for the maximum value of spin for each $l$ in figure.~\ref{figE} as an example.

\begin{figure}[tb]
\centering 
\includegraphics[width=.48\textwidth]{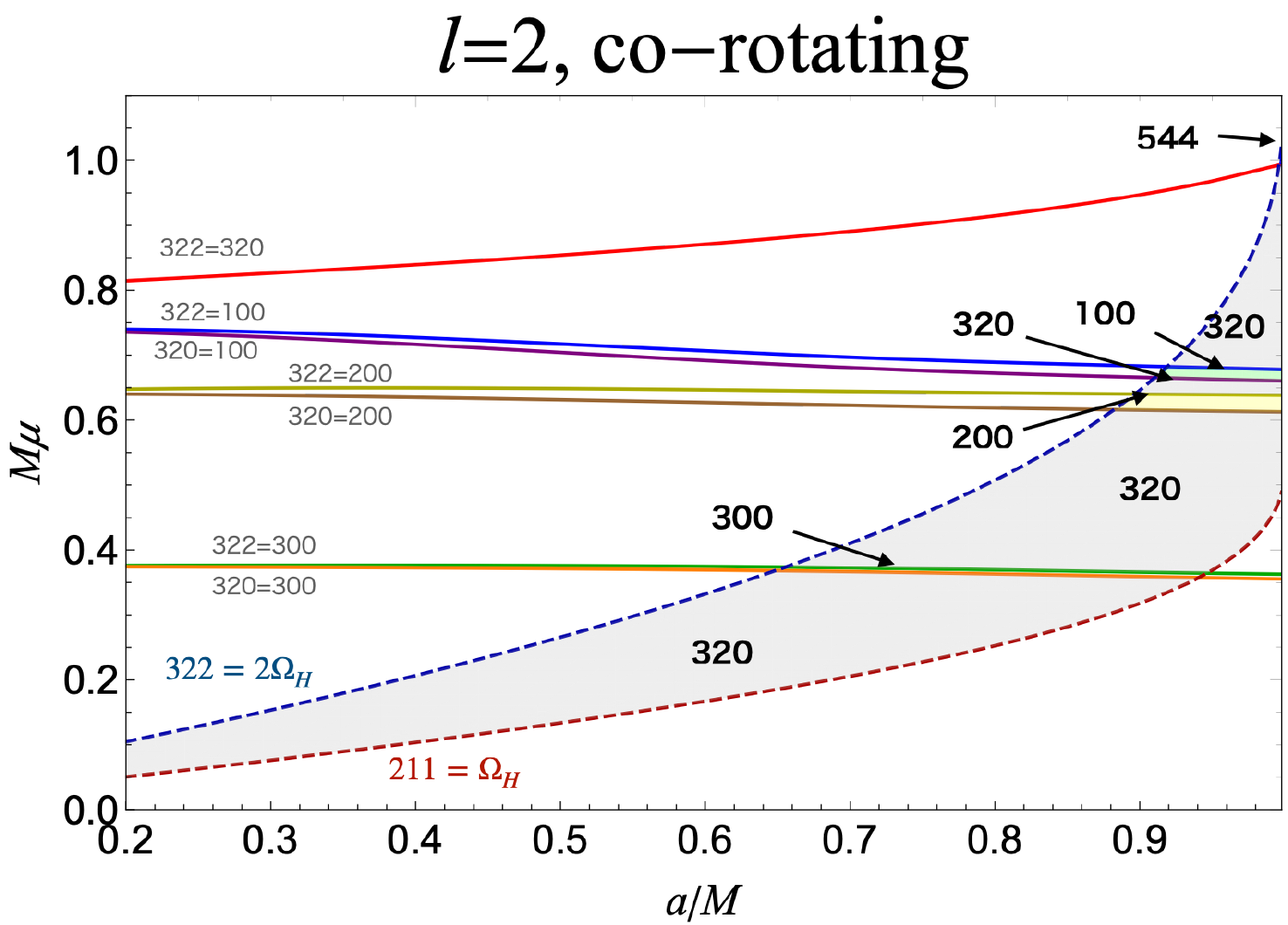}
\hfill
\includegraphics[width=.48\textwidth]{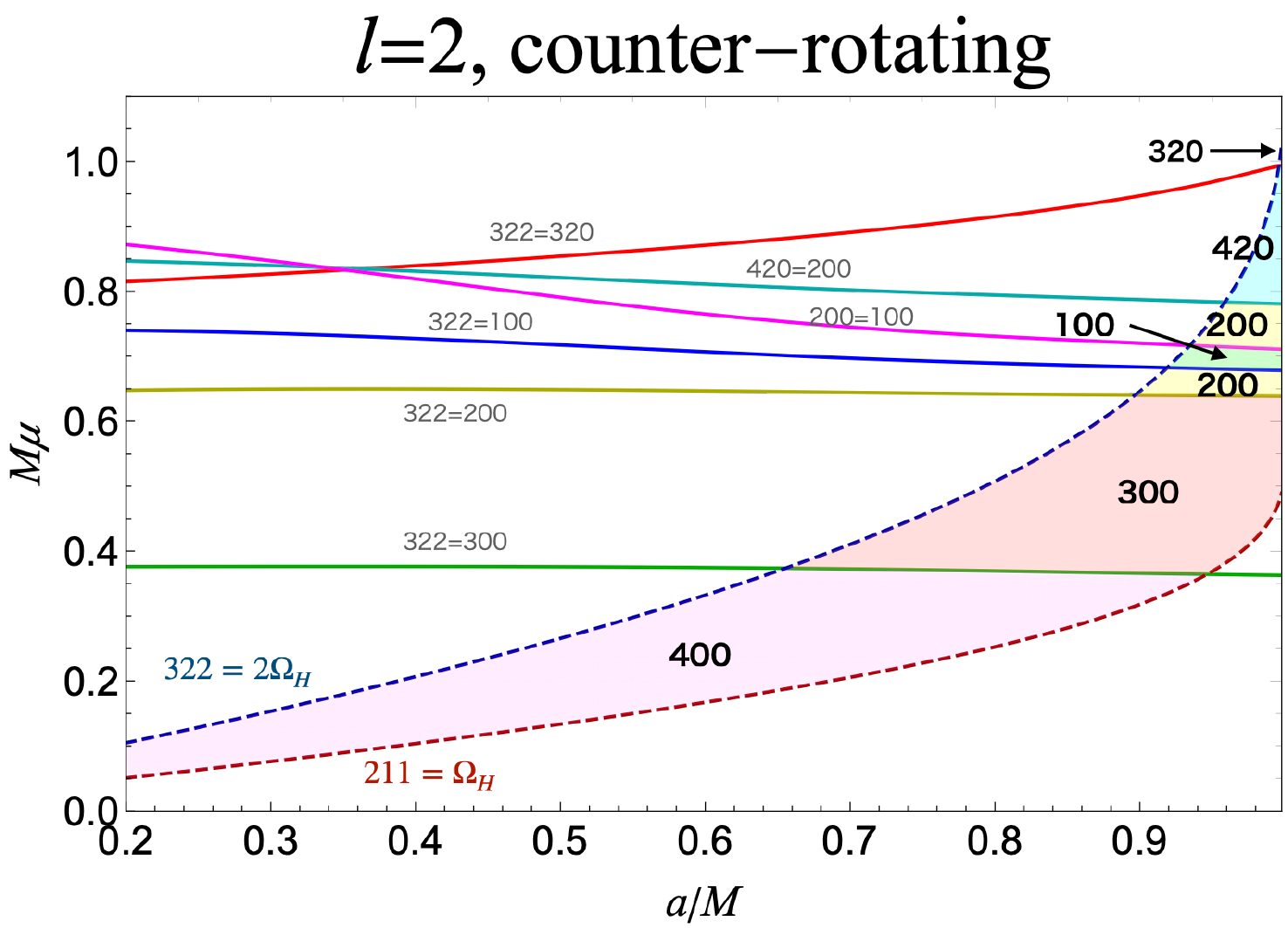} \\
\includegraphics[width=.96\textwidth]{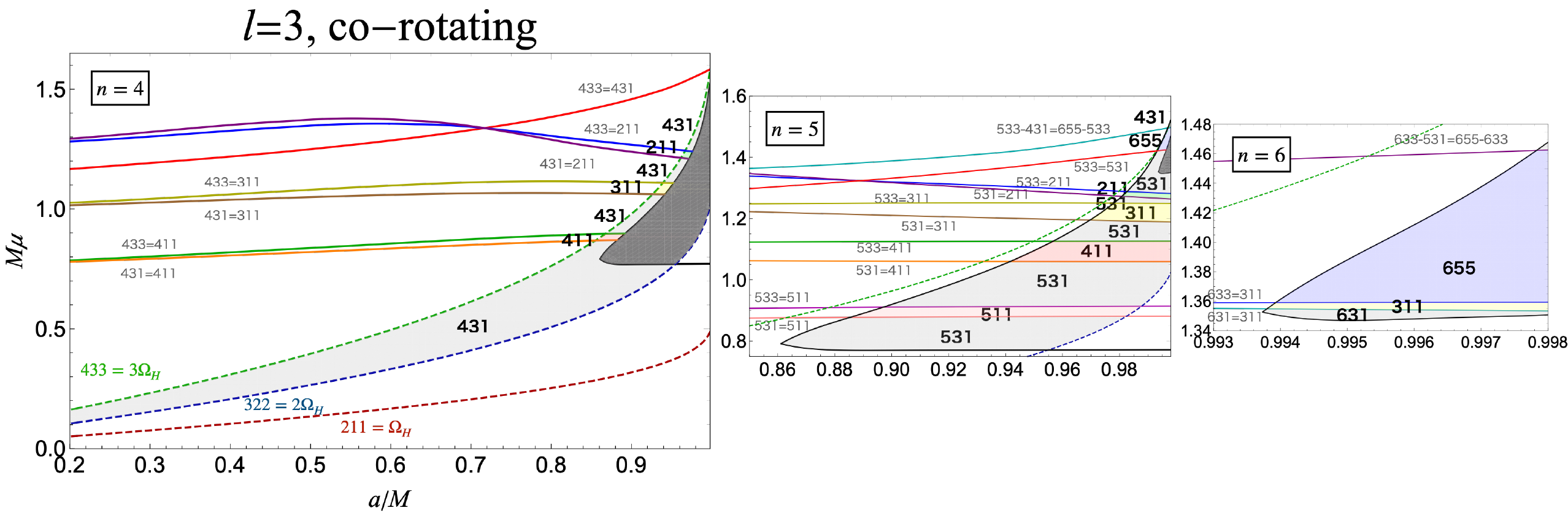}  \\
\includegraphics[width=.96\textwidth]{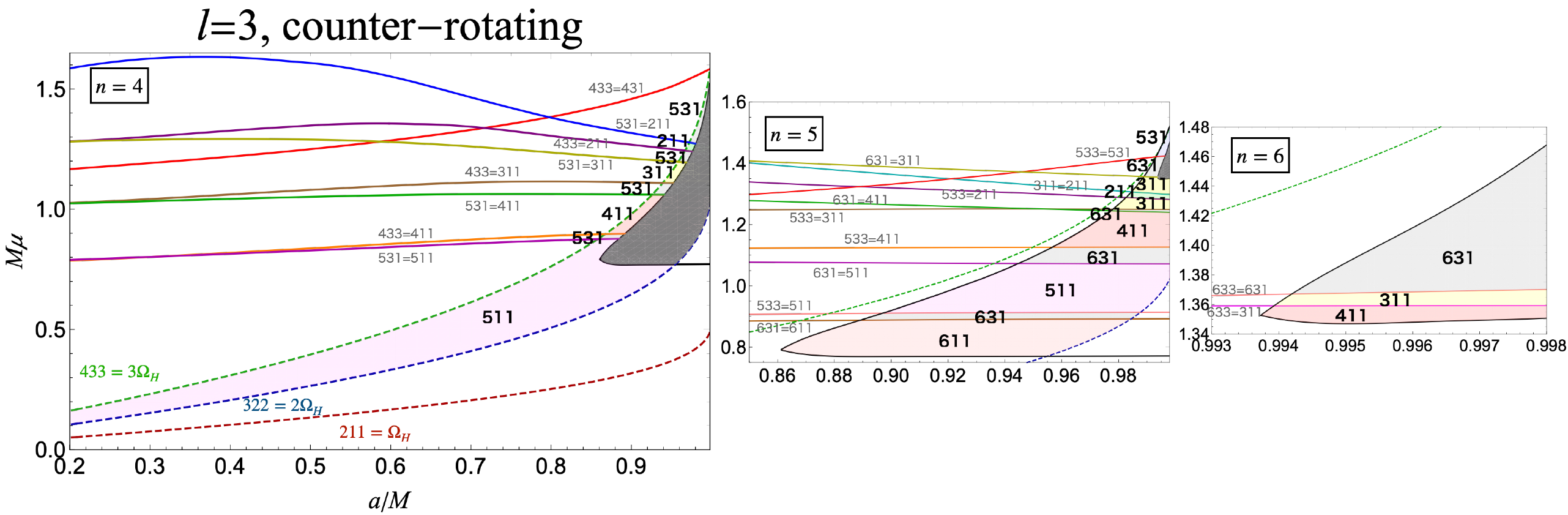} 
\caption{\label{figMap} Maps show the destination of the first transition of the axions occupying the fastest growing mode on the parameter plane. The top two panels represent the cases where $l=2$ mode is the fastest growing mode for the co-rotating case (left) and for the counter-rotating case (right), respectively. The middle three represent the case where $l=3$ mode is the fastest growing mode for the co-rotating case. The bottom three represent the case where $l=3$ mode is the fastest growing mode for the counter-rotating case. For $l=3$, the transition destination for each region is shown, when each $n$-th mode is the fastest growing mode, and the black shaded areas represent the region where the higher-$n$ mode grows faster. Dashed lines represent the boundaries for the superradiance condition to be satisfied for the representative values of $l$. Solid lines represent where the level crossing occurs. The colored area is the region of interest, and the number associated with each colored area indicates the transition destination.}
\end{figure}

From these figures, we can see that, because of the non-trivial behavior of energy levels, it is not possible to guess which transition occurs from the analogy of the hydrogen atom and it depends on the parameters. Comparing the energy differences from the focused growing mode, we classified the parameter space based on the destination of the first transition. The transition destination changes where the crossing between the relevant energy levels occurs. We made the map that shows the transition destination in the parameter plane as shown in figure.~\ref{figMap}. As a first step, we consider only quadrupolar tidal perturbation. For simplicity, we focus on the case in which the binary orbit is on the equatorial plane, i.e., only the transitions with $|\Delta m|=2$ are allowed. However, as we will see later, the conclusion will be unaltered even if we consider more general orbits, except for the case of an extremely large BH spin.

For $l=1$, level crossing between relevant modes does not occur in the superradiant region
\footnote{$(\omega_{R})_{211}$ and $(\omega_{R})_{21-1}$ coincide at $M\mu=4.97$, and the condition $(\omega_{R})_{211}<\Omega_{H}$ is violated at $M\mu=0.489$ for $a/M=0.998$.}, and hence the transition destinations are the same as the non-relativistic limit, i.e., $\ket{211}\to\ket{21-1}$ for the co-rotating case and $\ket{211}\to\ket{31-1}$ for the counter-rotating case. For $l=2$, as shown in figure.~\ref{figMap}, the transition destination depends on the values of parameters, but in almost the whole region, axions are transferred to a decaying mode which has the value of $m$ smaller than the original one. As an exception, in the co-rotating case, axions make a transition to the growing mode $\ket{544}$ in the region with a very large BH spin. However, in the presence of the superradiance, such a very large spin may not be realistic, since the process of the axion cloud formation extracts the angular momentum of the BH. For $l=3$, we should note that the growing mode with $n\neq l+1$ can be the fastest. Therefore, for each region where $n=4,5$ or $6$ is the fastest growing mode, we made a separate map showing the first transition destination. Similar to $l=2$, axions make a transition to a decaying mode in the almost whole region except for the region with a very large BH spin.

If the binary orbit is inclined, the transitions with $|\Delta m|=1$ are also allowed. Then, for example, $\ket{322}$ is first transferred to $\ket{321}$ in a wide parameter region for the co-rotating case. However, this means $(\omega_{R})_{322}-(\omega_{R})_{321}<(\omega_{R})_{321}-(\omega_{R})_{320}$, and hence the next transition $\ket{321}\to\ket{320}$ will occur successively. Therefore, the results described below should be almost unaltered. Furthermore, for the region with a very large BH spin where the transition to a growing mode occurs in the case of equatorial orbits, the transition to a decaying mode with $m$ less than one can occur in the non-equatorial case. On the other hand, in the counter-rotating case, the first transition destination will not change even if the orbit is inclined.

\subsection{Cloud depletion}
\begin{figure}[p]
\centering
\includegraphics[width=.47\textwidth]{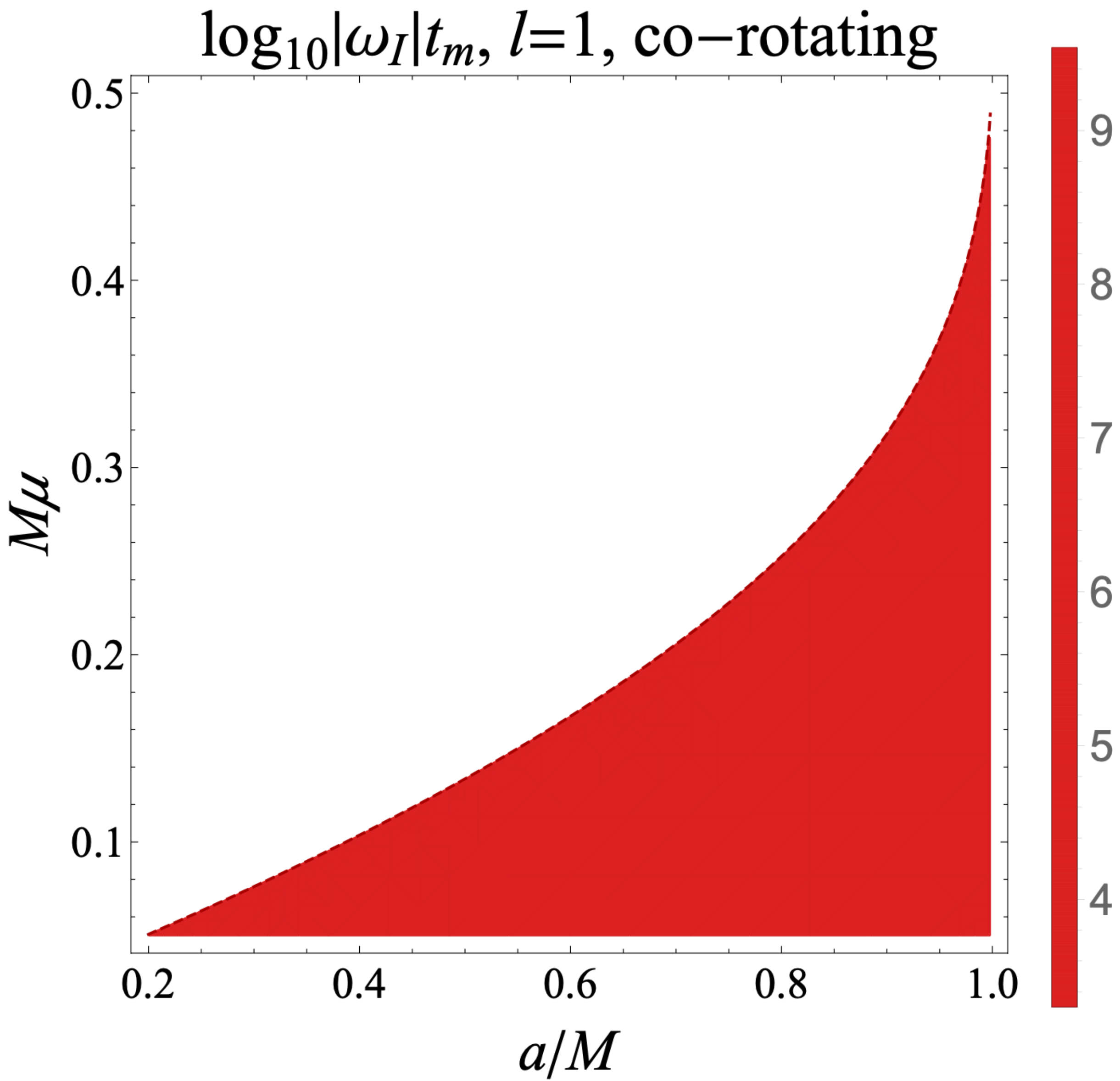}
\hfill
\includegraphics[width=.47\textwidth]{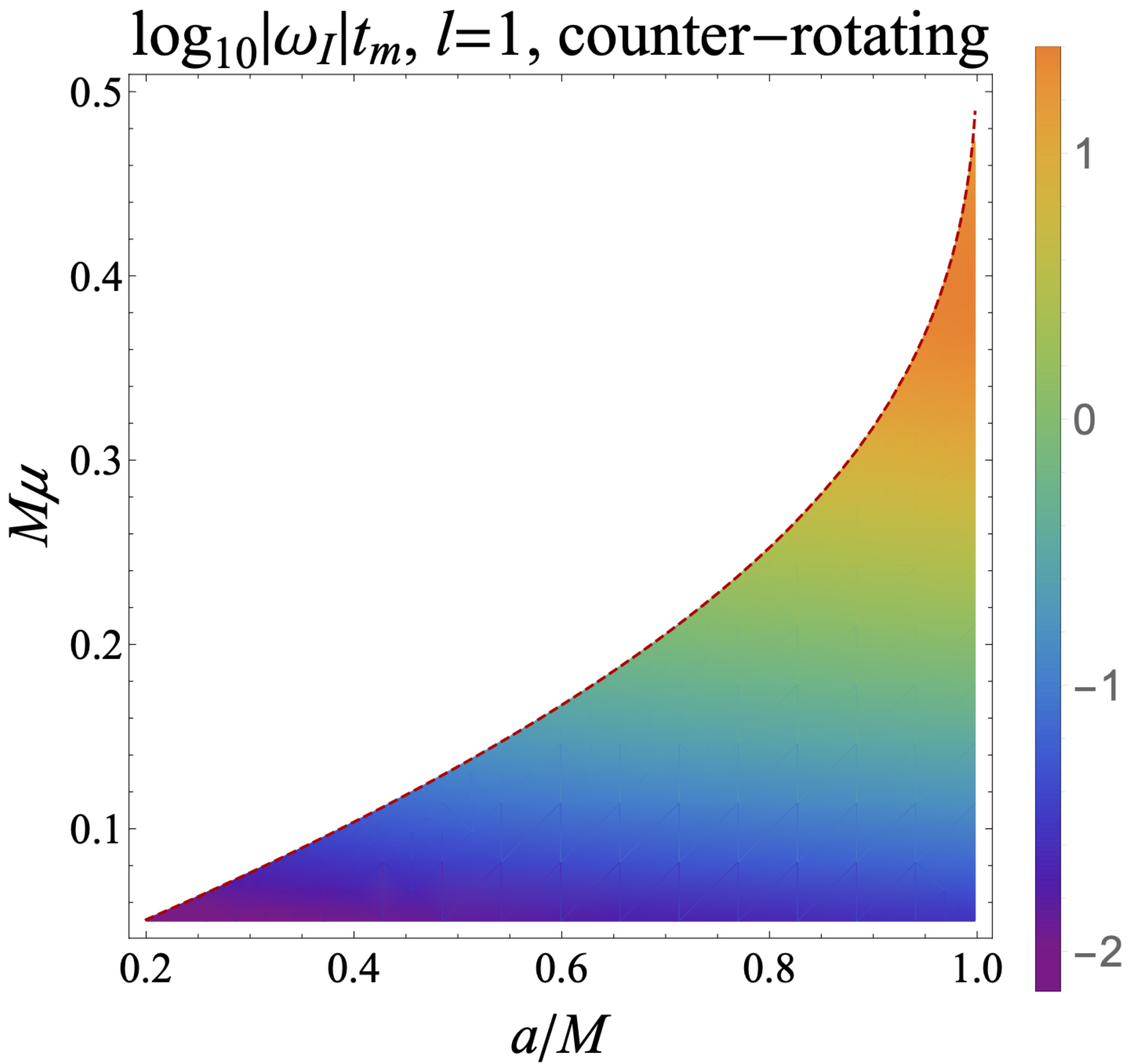} \\
\includegraphics[width=.47\textwidth]{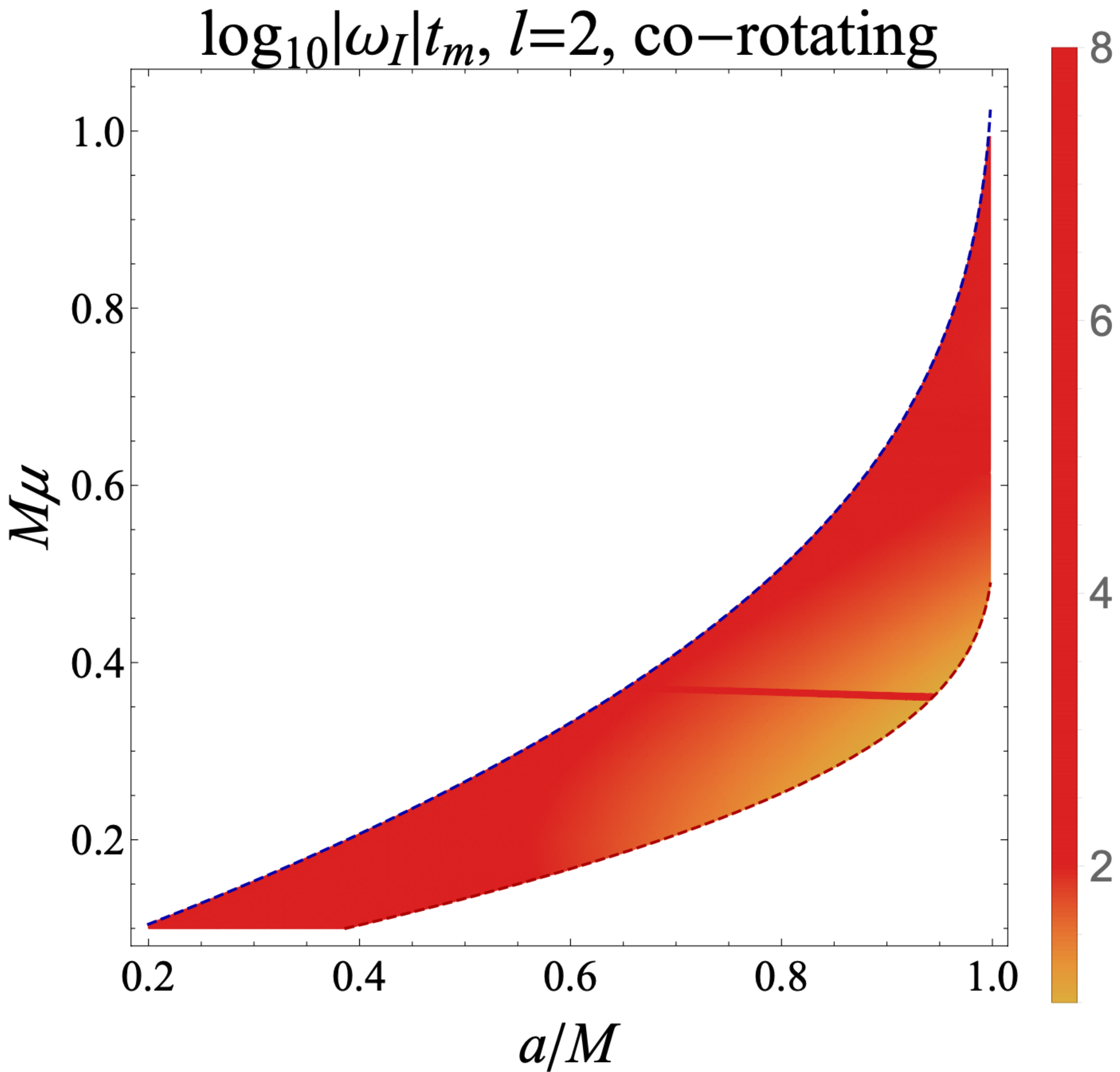}
\hfill
\includegraphics[width=.47\textwidth]{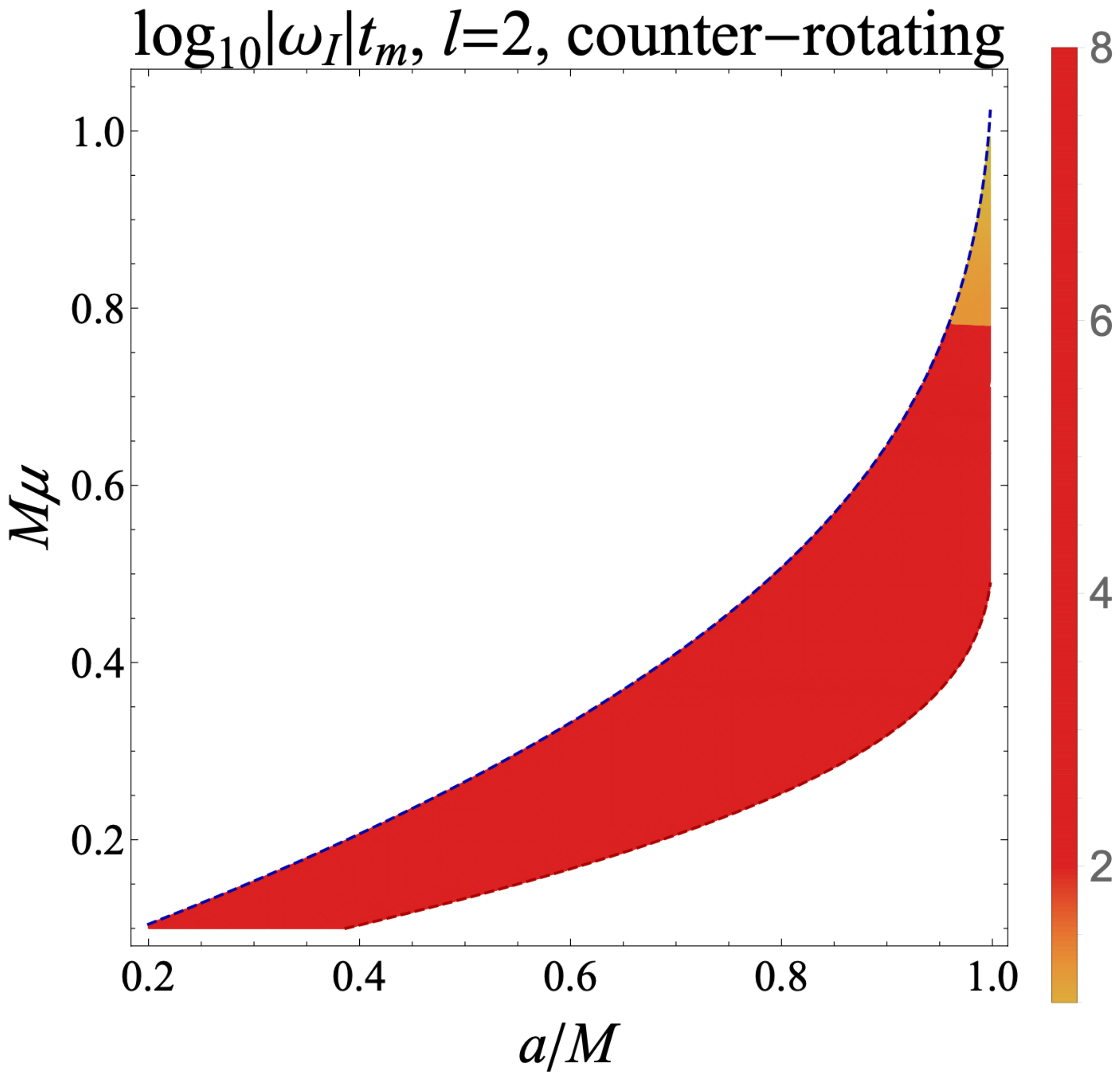} \\
\includegraphics[width=.47\textwidth]{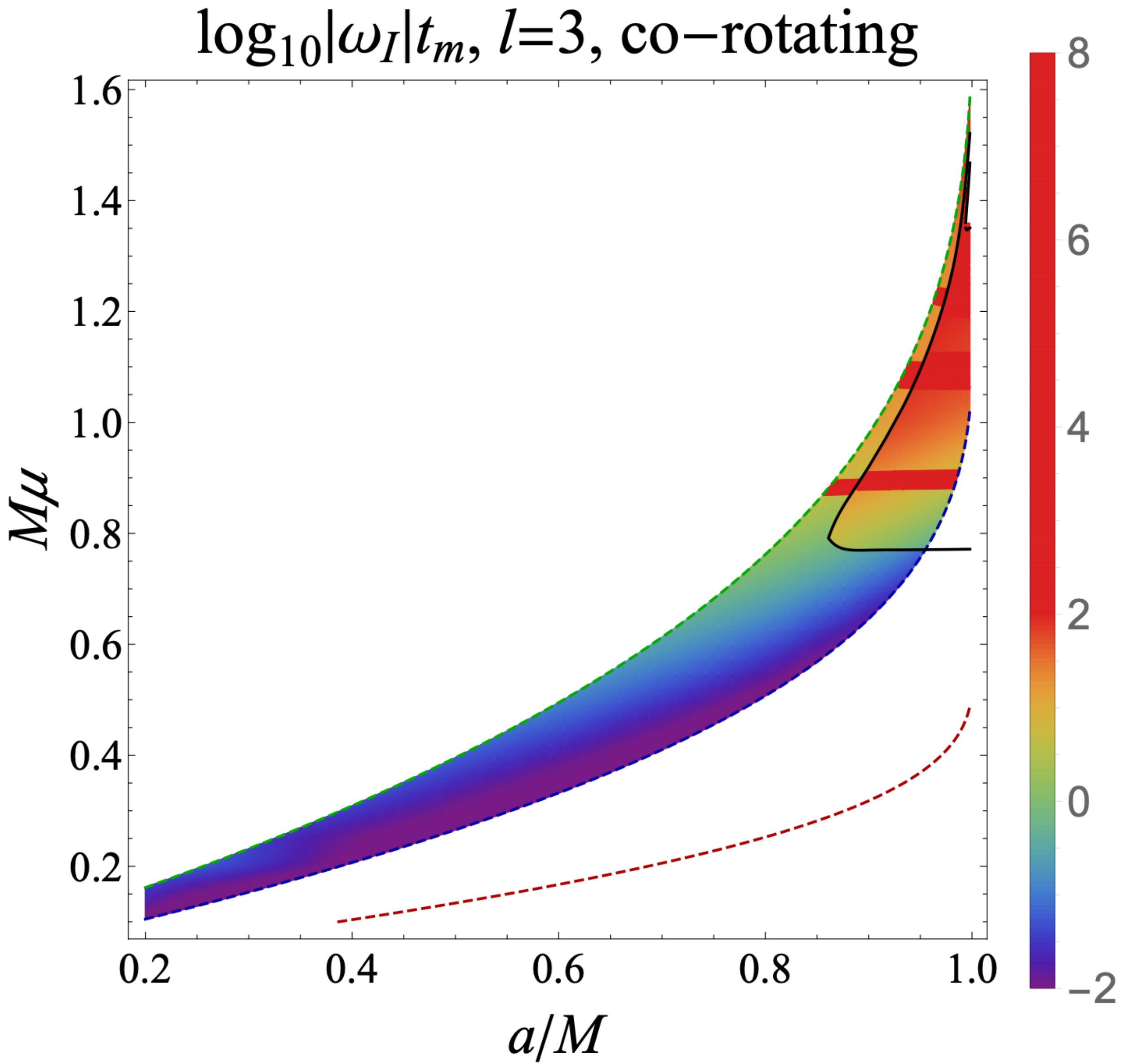}
\hfill
\includegraphics[width=.47\textwidth]{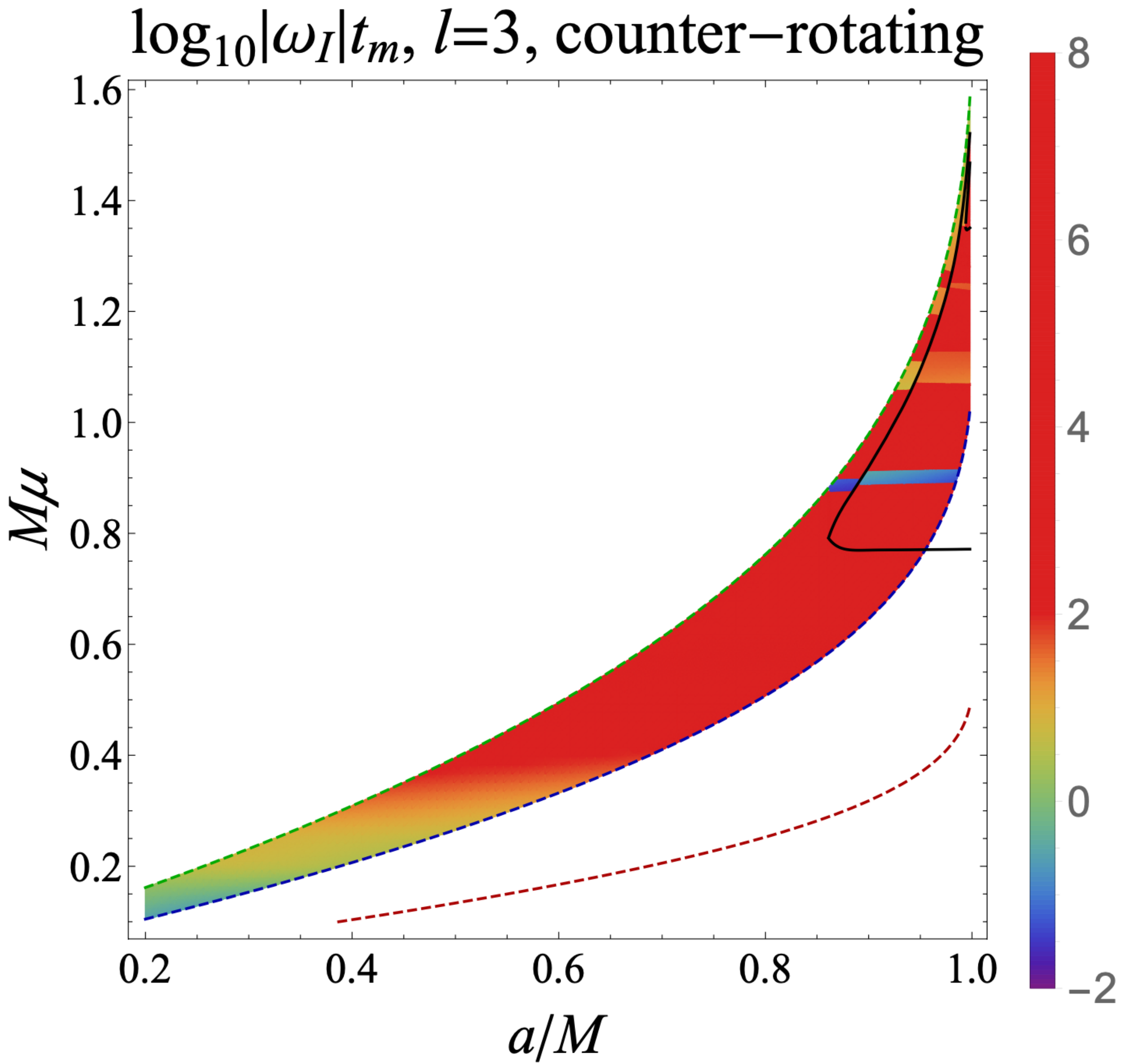}
\caption{\label{figExp} The plots show the value of $\log_{10}|\omega_{I}|t_m$, where $(\omega_{I})_{i}$ is the decay rate of the mode of the transition destination and $t_{m}$ is the time to elapse till the coalescence from the transition. The mass fraction of clouds that remain taking into account only the first transition is estimated by $\exp(-|\omega_{I}|t_m)$ for each case.  When $|\omega_{I}|_{i}t_{m}\gg1$ (the region in red), there is a sufficiently long time for the cloud to disappear before the merger phase. Note that we do not plot the narrow region where axions are first transferred to the growing mode.}
\end{figure}

Now that we know which transition occurs for each set of parameters, we can evaluate how much fraction of the cloud disappears between the first transition and the coalescence. Since the adiabatic condition of the transition is satisfied during the inspiral phase, we assume all axions are transferred to another mode at the resonance frequency immediately after crossing the resonance frequency. 
Then, we can estimate the mass of axion clouds after transition as
\footnote{As mentioned in Ref.~\cite{Baumann:2019ztm}, the estimation of the total amount of depletion in Ref.~\cite{Baumann:2018vus,Berti:2019wnn} assumes a constant orbital frequency. Therefore, they are not appropriate for the situation that we are considering.}
\begin{equation}
M_{cl}(t)\simeq M_{0}\exp\left[2(\omega_{I})t_{m}\right] \quad,
\end{equation}
where $M_{0}$ is the initial mass of axion clouds at the resonance frequency and $\omega_{I}$ is the imaginary part of the eigenfrequency of the transition destination. For quasi-circular orbits, the merger time taken from the resonance frequency to the coalescence is~\cite{PhysRev.136.B1224}
\begin{equation}
t_{m}=\frac{5}{256}\frac{(1+q)^{1/3}}{q}(M\Omega_{\mathrm{res}})^{-8/3}M \quad.
\end{equation}
To see whether or not axion clouds survive until around the merger phase, we should evaluate $|(\omega_{I})_{i}|t_{m}$, where $i$ is the label of the mode of transition destination for each parameter region shown in figure.~\ref{figMap}. In particular, since we focus on the observability of axion clouds with ground-based detectors, we consider the case with $q=1$. We show the numerical estimate about how a large fraction of the cloud mass disappears considering only the first transition in figure.~\ref{figExp}.

For $l=1$, in the co-rotating case, $|\omega_{I}|_{i}t_{m}$ is very large in the whole region. This means that the resonance frequency is so small, and the time before the coalescence is so long that there is a sufficiently long time for the clouds to disappear. Conversely, in the counter-rotating case, $|\omega_{I}|_{i}t_{m}$ is small, especially in the non-relativistic regime. This is because the resonance frequency is large in the counter-rotating case, and the time before the coalescence is short. However, for the relativistic regime, the decay rate is large, and hence $|\omega_{I}|_{i}t_{m}$ becomes large enough for the clouds to disappear.

For $l=2$, in the co-rotating case, $|\omega_{I}|_{i}t_{m}$ is very large in the whole region as in $l=1$ case. Also, in the counter-rotating case, $|\omega_{I}|_{i}t_{m}$ is very large in the whole region unlike in the $l=1$ case. Axions occupying the $l=2$ growing mode make a transition to the mode whose value of $l$ is smaller than the original one and its decay rate is very large because of the lower angular momentum barrier. Therefore, even with a short time before coalescence, the clouds almost disappear. Such transitions are absent in the $l=1$ case owing to the selection rule.

For $l=3$, in the region where transition destination is $\ket{n31} (n=4,5,6)$ except for the region with a large $M\mu$,  $|\omega_{I}|_{i}t_{m}$ is small because the decay rate of $\ket{n31}$ is so small. In the other region of interest,  $|\omega_{I}|_{i}t_{m}$ is so large that the clouds almost disappear.

\subsection{Subsequent transitions}
\begin{figure}[tb]
\centering
\includegraphics[width=.48\textwidth]{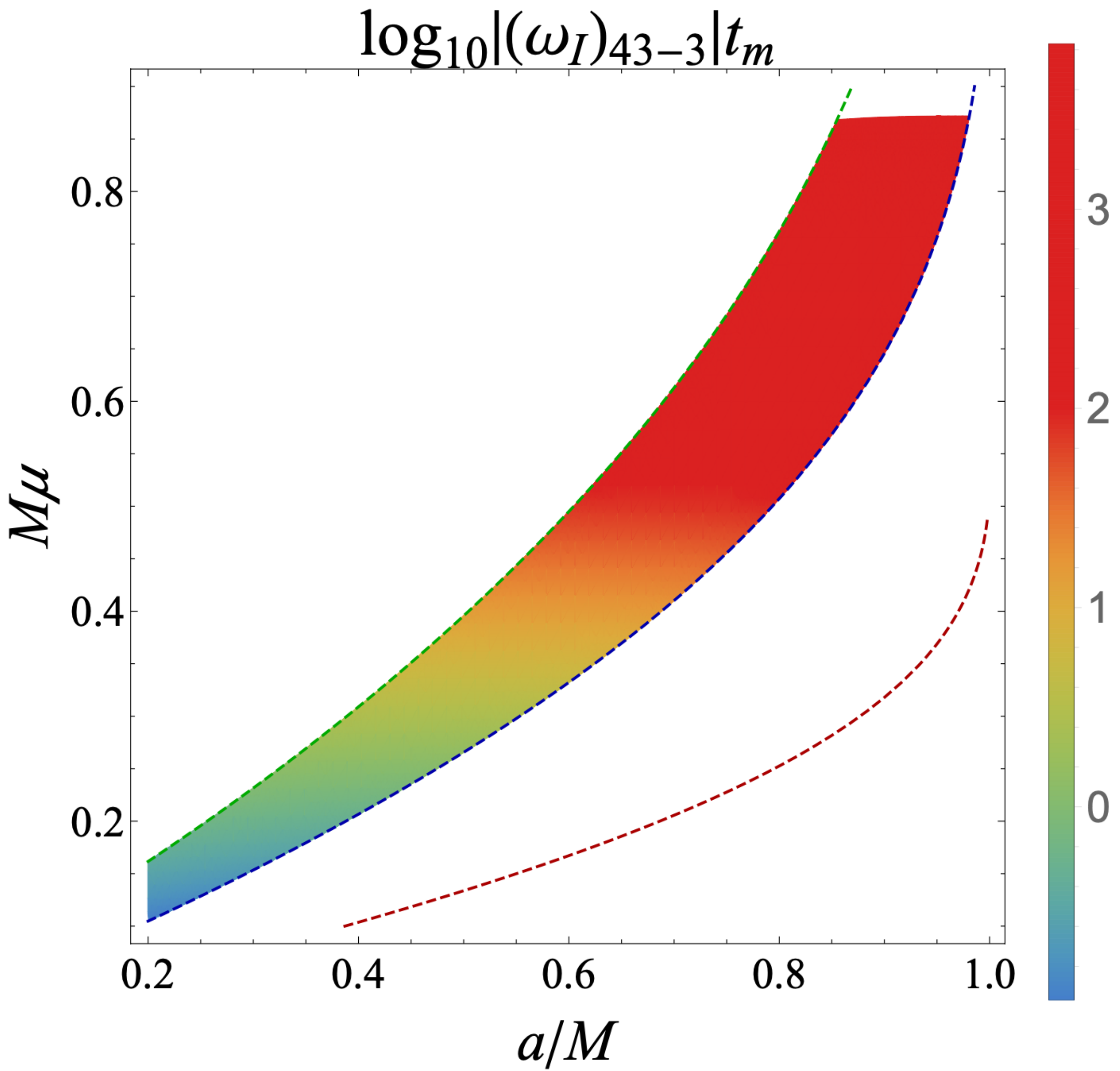}
\caption{\label{figExpSub} How much fraction of the clouds will disappear when axions that were occupying $\ket{433}$ are transferred to $\ket{43-3}$. These transitions occur in the co-rotating case.}
\end{figure}
As seen in figure.~\ref{figExp}, there are parameter regions where the clouds do not disappear if we consider only the first transition. To conclude whether the clouds in such parameter regions disappear or not, we need to investigate the subsequent transitions, too. 

For the $l=1$ counter-rotating case, axions are first transferred to $\ket{31-1}$. Since transitions occur only when $\Delta E/\Delta m <0$ and there is no mode with $l\geq1$ at energy levels lower than $\ket{31-1}$, axions will be next transferred to higher energy mode. Considering the orbits on the equatorial plane, i.e., allowing only transitions with $|\Delta m|=2$, because $(\omega_{R})_{31-1}-(\omega_{R})_{21-1}>\mu-(\omega_{R})_{31-1}$, the next transition will not occur. In the non-equatorial orbit case, axions can be next transferred to higher-$l$ mode, e.g. $\ket{53-2}$. However, they have a very small decay rate. Therefore, the cloud will not disappear at least when we take into account only the quadrupolar tidal perturbation.

For the $l=3$ co-rotating case, considering the transition $\ket{433}\to\ket{431}$, successive transitions $\ket{431}\to\ket{43-1}\to\ket{43-3}$ will occur. Hence, we should judge whether the clouds disappear or not by evaluating $|(\omega_{I})_{43-3}|t_m$. The result is shown in figure.~\ref{figExpSub} and it is large at $M\mu\gtrsim0.3$. Notice that the original growth rate of higher-$l$ mode is suppressed by the higher angular momentum barrier and there is less physical interest in the small $M\mu$ region for $l=3$. A similar thing happens to $n=5$. In the counter-rotating case, axions that experience the transition to $\ket{531}$ or $\ket{631}$ will then be transferred to the mode with $m=-1$. Its decay rate should be very large and the clouds disappear.

\section{Conclusion and Discussion}\label{conclusion}
In this paper, we focused on the observability of axion clouds in the waveform of GWs from coalescing BH binaries. If the clouds are still present in the late inspiral phase, gravitational waveform observed with ground-based detectors can also be affected. Considering also the case when the cloud is in the relativistic regime, whether or not the tidal interaction depletes the cloud is not trivial. As a first step of the exhaustive study of cloud depletion, we investigated the quadrupolar tidal effect in the parameter region where the growth rate of the mode with $l=1,2$ or $3$ is the fastest.

First, we confirmed that the axions occupying the growing mode are transferred to a decaying mode first in almost the whole parameter region. The exception is when the BH spin is extremely large, and the binary orbit is restricted to the equatorial plane with respect to the black hole spin. Then, assuming that the transition is adiabatic, we evaluated whether or not there is a sufficiently long time for clouds to disappear between the transition and the coalescence for equal mass binaries. The result is that the quadrupolar tidal perturbation does not deplete the cloud only when $l=1$ mode is the fastest growing mode, the binary orbit is counter-rotating, and the cloud is in the non-relativistic regime.  In other cases, the cloud will deplete well in advance before the merger phase.

Since the phenomenon we are considering is a resonant effect, even a higher multipole moment would also induce transition, if it is sufficiently large and the resonant frequency is hit earlier. As mentioned in Ref.~\cite{Baumann:2019ztm}, considering the octupolar tidal perturbation, the cloud also depletes when $l=1$ mode is the fastest and the orbit is counter-rotating. However, it is not clear that how higher multipole moments affect the transition network especially in the region where the non-trivial transition occurs. In addition, when the cloud is in the relativistic regime, the configuration is largely different from the one of the hydrogen atom. Hence, the overlap between mode functions will change and whether the adiabatic condition is satisfied or not is non-trivial. Therefore, we should solve the transition network taking into account the higher multipole moments and evaluate the transition rate appropriately for each transition in order to get a conclusive answer to the consequence of perturbative tidal effects. Our work is the first step in the investigation of cloud depletion in a wide parameter region.

Furthermore, if the condition under which the cloud can evade the depletion is found, we need to study the evolution of the cloud after the binary companion enters the cloud. In such a case, the previously developed perturbative analysis cannot be applied, and a completely different treatment will be necessary. There may also be another scenario in which clouds do not disappear. For example, it is considered that a companion might be directly captured by a BH without passing the resonance frequency. We leave these problems as future work.

\acknowledgments
We would like to thank H.~Omiya for the fruitful discussions. This work is supported by JSPS Grant-in-Aid for Scientific Research JP17H06358 (and also JP17H06357), as a part of the innovative research area, ``Gravitational wave physics and astronomy: Genesis'', and also by JP20K03928. 

\appendix
\section{Tidal perturbation and Selection rules}\label{AppTidal}
Following Ref.~\cite{Baumann:2018vus}, we summarize the tidal perturbation from the binary companion and the selection rule of level transitions. Let $\bm{r}(t)=(r(t),\theta(t),\varphi(t))$ be the comoving Fermi coordinates with the center of mass of the BH-cloud, and $\bm{R}_{\ast}(t)=(R_{\ast}(t),\Theta_{\ast}(t),\Phi_{\ast}(t))$ be the coordinates of the binary companion. At Newtonian order, the tidal potential can be written as
\begin{equation}
V_{\ast}(t)=-\frac{M_{\ast}\mu}{R_{\ast}}\sum_{l_{\ast}=2}^{\infty}\sum_{|m_{\ast}|\leq l_{\ast}}\frac{4\pi}{2l_{\ast}+1}\left(\frac{r}{R_{\ast}}\right)^{l_{\ast}}Y_{l_{\ast}m{\ast}}^{\ast}(\Theta_{\ast},\Phi_{\ast})Y_{l_{\ast}m_{\ast}}(\theta,\varphi) \quad ,
\end{equation}
where $Y_{lm}$ are the spherical harmonics. For non-relativistic limit, axion's eigenstates are given by $\psi_{i}=e^{-i(\omega-\mu)t}R_{nl}(r)Y_{lm}(\theta,\varphi)$. The perturbation to the dynamics of an axion cloud, i.e. the overlap between different two modes are
\begin{equation}
\braket{i'|V_{\ast}|i}=-\frac{M_{\ast}\mu}{R_{\ast}}\sum_{l_{\ast}=2}^{\infty}\sum_{|m_{\ast}|\leq l_{\ast}}\frac{4\pi}{2l_{\ast}+1}\frac{Y^{\ast}_{l_{\ast}m_{\ast}}(\Theta_{\ast},\Phi_{\ast})}{R_{\ast}^{l_{\ast}}}\times I_{r}\times I_{\Omega} \quad ,
\end{equation}
where
\begin{align}
I_{r}&=\int dr\ r^{2+l_{\ast}}R_{n'l'}(r)R_{nl}(r) \quad ,\\
I_{\Omega}&=\int d\Omega\ Y^{\ast}_{l'm'}(\theta,\varphi)Y_{l_{\ast}m_{\ast}}(\theta,\varphi)Y_{lm}(\theta,\varphi) \quad .
\end{align}
From the angular integral, the overlap does not vanish if and only if the following selection rules are satisfied:
\begin{itemize}
\item $-m'+m_{\ast}+m=0\quad,$ 
\item $|l-l'|\leq l_{\ast}\leq l+l'\quad,$
\item $l'+l_{\ast}+l=2p, ~(p\in \mathbb{Z})\quad.$
\end{itemize}
In particular, we focus on the leading quadrupolar perturbation $l_{\ast}=2$. 

\section{Continued Fraction Method}\label{AppCF}
We calculated the eigenfrequencies of axion's bound states numerically using a continued fraction method~\cite{doi:10.1063/1.527130,Dolan:2007mj}. The equation of motion Eq.~(\ref{eom}) can be decomposed into radial and angular equations as
\begin{equation}\label{SphEOM}
\frac{1}{\sin\theta}\frac{\rm{d}}{\rm{d}\theta}\left(\sin\theta\frac{{\rm{d}}
S_{lm\omega}}{\rm{d}\theta}\right)
+\left[c^2(\omega)\cos^2\theta-\frac{m^2}{\sin^2\theta}\right]S_{lm\omega}
=-\Lambda_{lm}(\omega)S_{lm\omega} \quad,
\end{equation}
\begin{align}
\frac{\rm{d}}{{\rm{d}}r}
\left(\Delta\frac{{\rm{d}}R_{lm\omega}}{{\rm{d}}r}\right)
+&\Biggl[\frac{(r^2+a^2)^2\omega^2-4amM\omega r+a^2m^2}{\Delta}  \notag \\
&\quad -(a^2\omega^2+\mu^2r^2+\Lambda_{lm}(\omega))\Biggr]R_{lm\omega}=0 \label{radial} \quad,
\end{align}
where $\Delta=r^2-2Mr+a^2$ and $c^2(\omega)=a^2(\omega^2-\mu^2)$.  Usually, the angular eigenvalue $\Lambda_{lm}$ is calculated by power series expansion of $c$ around that of spherical harmonics $l(l+1)$~\cite{Berti:2005gp}. However, we are also interested in the region where $c\ll1$ is not satisfied. Then, we calculated $\Lambda_{lm}$ exactly by continued fraction method, and summarize it here. 

Imposing the regularity at $u=\pm1$, the angular mode function can be written as
\begin{equation}
S_{lm\omega}=(1+u)^{|m|/2}(1-u)^{|m|/2}e^{i(-ic)u}\sum_{n=0}^{\infty}a_{n}^{\theta}u^{n} \quad,
\end{equation}
where $u=\cos\theta$. From Eq.~(\ref{SphEOM}), the expansion coefficients $a_{n}^{\theta}$ satisfy the following three-trem recurrence relation:
\begin{align}
&\alpha_{0}^{\theta}a_{1}^{\theta}+\beta_{0}^{\theta}a_{1}^{\theta}=0 \quad,\\
&\alpha_{n}^{\theta}a_{n+1}^{\theta}+\beta_{n}^{\theta}a_{n}^{\theta}+\gamma_{n}^{\theta}a_{n-1}^{\theta}=0, \quad (n=1,2,\cdots) \quad, \label{three}
\end{align}
where
\begin{align}
\alpha_{n}^{\theta}&=-2(n+1)(n+|m|+1) \quad, \\
\beta_{n}^{\theta}&=n(n-1)+2n(|m|+1-2c)-(2c(|m|+1)-|m|(|m|+1))-(c^2+\Lambda_{lm}) \quad, \\
\gamma_{n}^{\theta}&=2c(n+|m|) \quad.
\end{align}
The ratio of successive elements of the convergent solution sequence is given by the continued fraction
\begin{equation}
\frac{a_{n+1}^{\theta}}{a_{n}^{\theta}}=-\frac{\gamma_{n+1}^{\theta}}{\beta_{n+1}^{\theta}-}\frac{\alpha_{n+1}^{\theta}\gamma_{n+2}^{\theta}}{\beta_{n+2}^{\theta}-}\frac{\alpha_{n+2}^{\theta}\gamma_{n+3}^{\theta}}{\beta_{n+3}^{\theta}-}\cdots
\end{equation}
and $a_{1}^{\theta}/a_{0}^{\theta}=-\beta_{0}^{\theta}/\alpha_{0}^{\theta}$. Note that in the limit of $c\to0$, $\Lambda_{lm}$ should become $l(l+1)$. In this limit, $\gamma_{n}^{\theta}=0$ for all $n$, and the sequence stops at $n$ such that $\beta_{n}^{\theta}=0$. This will happen when $\Lambda_{lm}=(n+|m|)(n+|m|+1)$, that is, when $n=l-|m|$. Therefore, setting $n=l-|m|$ in Eq.~(\ref{three}), we obtain
\begin{align}
-\frac{\alpha_{l-|m|}^{\theta}\gamma_{l-|m|+1}^{\theta}}{\beta_{l-|m|+1}^{\theta}-}&\frac{\alpha_{l-|m|+1}^{\theta}\gamma_{l-|m|+2}^{\theta}}{\beta_{l-|m|+2}^{\theta}-}\cdots+\beta_{l-|m|}^{\theta} \notag \\
&+\frac{\alpha_{l-|m|-1}^{\theta}\gamma_{l-|m|}^{\theta}}{\beta_{l-|m|-1}^{\theta}-}\frac{\alpha_{l-|m|-2}^{\theta}\gamma_{l-|m|-1}^{\theta}}{\beta_{l-|m|-2}^{\theta}-}\cdots\frac{\alpha_{0}^{\theta}\gamma_{1}^{\theta}}{\beta_{0}^{\theta}}=0 \quad.
\end{align}
Solving this equation numerically, we can obtain the angular eigenvalue $\Lambda_{lm}$. Similarly, we can calculate the radial eigenvalue, i.e., eigenfrequency. See Ref.~\cite{Dolan:2007mj} for details.

\bibliography{ref}
\bibliographystyle{JHEP}




\end{document}